\documentclass[a4paper,11pt]{article}
\usepackage{pos}
\usepackage{svg}

\title{Dark Matter searches in Dwarf Galaxies with the Southern Wide-field Gamma-ray Observatory}
 \ShortTitle{DM searches in Dwarf Galaxies with the SWGO}

\author[a,*]{Micael Andrade\footnotetext{$^*$Speaker}}
\author[a]{Aion Viana}

\affiliation[a]{Instituto de Física de São Carlos, Universidade de São Paulo\\
  Av. Trabalhador São Carlense 400, São Carlos, Brazil}

\onbehalf{for the SWGO collaboration$^\dagger$\footnotetext{$^\dagger$For the complete list of the SWGO Collaboration and acknowledgments, please visit: \url{https://www.swgo.org/SWGOWiki/lib/exe/fetch.php?media=wiki:swgo_collaboration_icrc23.pdf}}} 

\emailAdd{micaelandrade@ifsc.usp.br}
\emailAdd{aion.viana@ifsc.usp.br}

\abstract{Dark matter is thought to make up most of the matter density of the Universe, yet its true nature remains uncertain. Among dark matter theories, Weakly Interacting Massive Particles (WIMPs) are a prominent candidate for dark matter because they can reproduce the observed abundance of dark matter in the universe. There are various methods for searching for WIMPs, one of which is indirect detection, which involves looking for the Standard Model particles produced by the decay or self-annihilation of dark matter particles. Within the mass range of GeV to PeV for the dark matter particle, this type of search can be conducted by detecting $\gamma$-rays in astrophysical objects with high concentrations of dark matter. Dwarf galaxies, although not the most dense, are excellent targets for this type of observation since they are dominated by dark matter, are relatively close to Earth, and have a low astrophysical background. In this work, the detectability of dark matter annihilation or decay signals from dwarf galaxies is predicted using the Southern Wide-field Gamma-ray Observatory (SWGO), a future $\gamma$-ray observatory that will be built in South America. This wide field-of-view survey instrument will be able to study many important dark matter targets in the Southern Hemisphere, and the combined observation of all targets will provide competitive, if not the best, limits for dark matter with masses in the range of hundreds of GeV to PeV.}

\ConferenceLogo{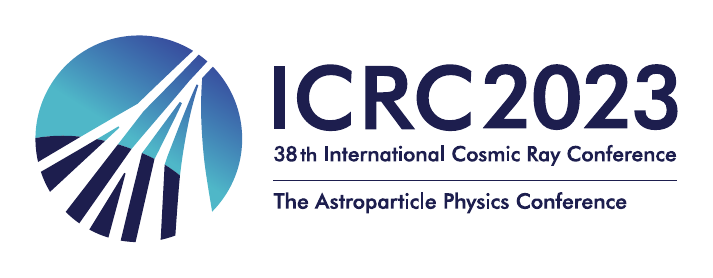}

\FullConference{%
38th International Cosmic Ray Conference (ICRC2023)\\
  26 July - 3 August, 2023\\
  Nagoya, Japan}

\begin{document}
\maketitle

\section{Introduction}

The mystery surrounding dark matter remains one of the most profound mysteries in contemporary astrophysics, ever since Fritz Zwicky first identified the mass-to-light disparity in the Coma Cluster~\cite{1933AcHPh...6..110Z}. Over time, a growing amount of evidence, including rotation curves, gravitational lensing, and the formation of large structures, have been supporting the existence of dark matter~\cite{1970ApJ...159..379R, 2006ApJ...648L.109C, 2006Natur.440.1137S}. Consequently, theoretical physicists have proposed a plethora of particles that could explain dark matter in attempts to elucidate its true essence \cite{2005PhR...405..279B}. Among them, WIMPs (weakly interacting massive particles) are one of the most compelling candidates, as they have a mass in the GeV to TeV range, and naturally reproduce the observed DM abundance, if they have an annihilation cross-section around the weak scale \cite{Arcadi:2017kky}. Nevertheless, the enigmatic nature of dark matter persists, despite the amount of effort to probe this elusive particle.

Indirect detection aims at observing the secondary products from the interactions of dark matter particles, such as high-energy photons, neutrinos, and cosmic rays. Through the study of dark matter dominated systems in the Universe, we are able to restrict the parameter space of the properties of different dark matter candidates. These dark matter dominated systems are expected to be prone to decay and self-annihilation of dark matter particles and within the GeV to TeV for the mass of the dark matter particle, it is expected a significant emission of gamma-rays from these processes~\cite{2005MPLA...20.1021B}.

The Southern Wide-field Gamma-ray Observatory (SWGO) is a future observatory to be constructed in South America and will consist of a ground-based air shower detector array. SWGO will be sensitive to gamma-rays with energies from 100 GeV to hundreds of TeV \cite{2019arXiv190208429A}. The main focus of this work is to investigate the capabilities of SWGO to constrain the WIMPs parameter space with observations of dwarf galaxies. 

\section{Dark matter search in dwarf galaxies}

The exploration of Dark Matter through indirect methods presents numerous potential targets, including the galactic center, dwarf galaxies, galaxy clusters, and even the Sun. Each of these targets possesses its own advantages and disadvantages. For instance, the galactic center holds the promise of yielding the most significant signal of gamma-rays stemming from dark matter annihilation or decay. Nevertheless, the galactic center also presents the most intricate background, which poses considerable challenges in detecting a dark matter signal~\cite{2005MPLA...20.1021B} . 

There are two usual signals searched in indirect detection of dark matter, annihilation and decay. The flux of photons from the annihilation of dark matter particles is given by

\begin{equation}
         \textsc{Gamma-ray flux} = \overbrace{\frac{\langle \sigma v \rangle}{8\pi m_{DM}^2} \frac{dN}{dE}}^{\textsc{particle physics} }\underbrace{\int ds\int d\Omega\,\; \rho_{DM}^2}_{\textsc{J-factor}} \, ,
         \label{eqn:fluxann}
\end{equation}
where $\langle \sigma v \rangle$ is the annihilation cross section, $m_{DM}$ is the mass of the dark matter particle, and $\frac{dN}{dE}$ is the annihilation spectrum.

Similarly, the formula for the flux coming from decaying dark matter particles with mass $m_{DM}$:
\begin{equation}
         \textsc{Gamma-ray flux} = \overbrace{\frac{1}{4\pi \tau m_{DM}} \frac{dN}{dE}}^{\textsc{particle physics}}\underbrace{\int ds\int d\Omega\,\; \rho_{DM}}_{\textsc{D-factor}} \, ,
         \label{eqn:fluxdec}
\end{equation}
where $\tau$ is the decay time.

These gamma-ray flux equations involve two components, one which is dictated by particle physics and another one which is dictated by astrophysics and it is usually called J-factor for annihilation or D-factor for decay. The density profile is one of the most important features of dark matter models, consequently, the J-factor and D-factor are also extremely important. As these astrophysical factors are integrated over the line of sight and solid angle, the closer the target is to Earth the higher their astrophysical factor would become. Hence, the distance from the target plays a crucial role in determining the optimal targets. 

Besides these factors, it is also important to consider the annihilation or decay spectrum into various individual channels. These channels yield Standard Model particles through annihilation or decay. This work considers the spectrum calculated in PPPC 4 DM ID \cite{2011JCAP...03..051C}.

The choice of target is also critical. The Local Group has several dwarf galaxies which are satellites galaxies of the Milky Way, implying a proximity to Earth. Among these dwarf galaxies, dwarf spheroidal galaxies (dSphs) assume particular prominence for indirect searches of dark matter due to their distinctive characteristics. dSphs possess an older stellar population and lack active star formation, resulting in minimal gamma-ray emission if dark matter does not exist. These environments, devoid of astrophysical contaminants, provide the cleanest possible dark matter signature compared to other targets. Furthermore, dSphs are dominated by dark matter, allowing for the measurement of density profiles based on stellar dynamics.

\begin{figure}
 \centering
 \includegraphics[width=\textwidth]{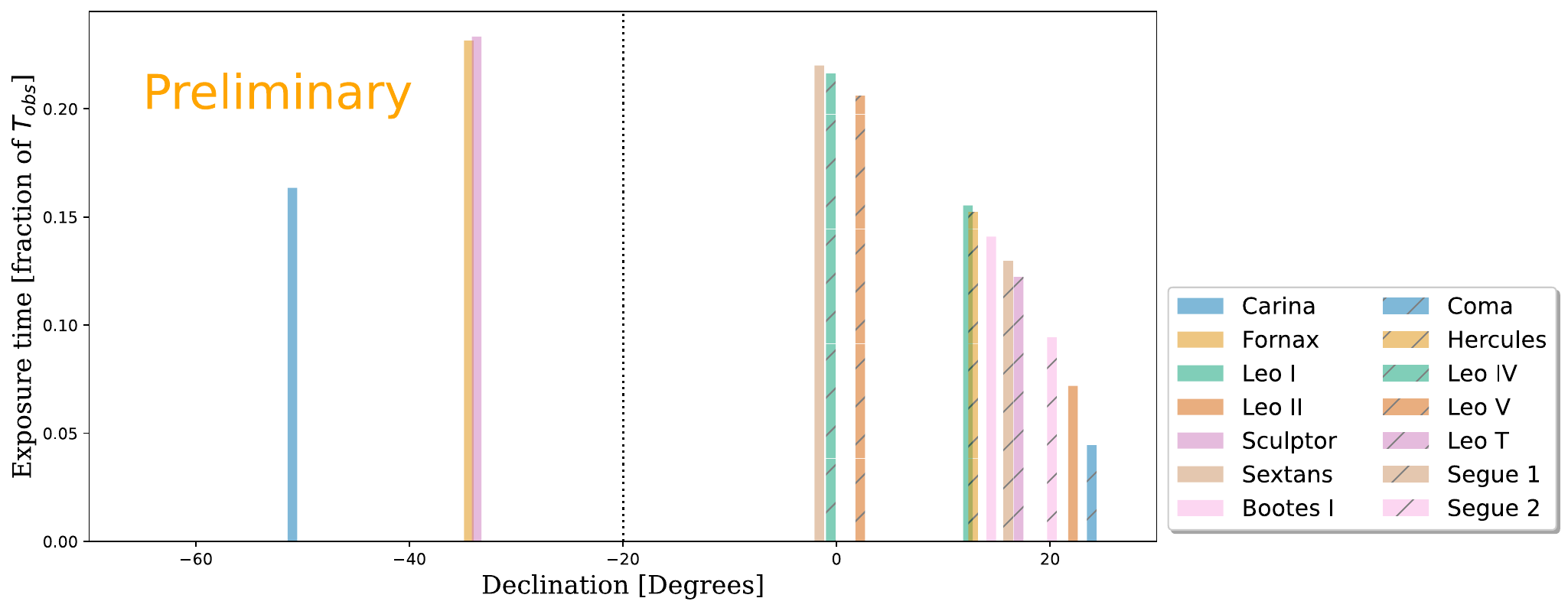}
 \caption{Exposure time as a function of declination for each Dwarf galaxy (in Ref.~\cite{2015ApJ...801...74G}) in the field of view of the SWGO. This considers a latitude of SWGO at -20$^{\circ}$ and a maximum zenith angle of observation of 45$^{\circ}$. Dotted line represents SWGO position and $T_{obs}$ is the total observation time with SWGO.}
 \label{fig_exposuretime}
\end{figure}

The dwarf galaxies in the range of the SWGO depends on its localization in the sky. In this work, it was considered a latitude of -20 degrees for the observatory site (in the range the current candidate sites) and a maximum zenith angle of 45 degrees, based on current capabilities of HAWC. It is important to assess how this specific position affects the dwarf galaxies that are within the field of view of the SWGO. Figure \ref{fig_exposuretime} presents dwarf galaxies within the SWGO range at this position, along with their respective exposure times. This work considered the dwarf galaxies and their properties as derived by Geringer-Sameth \emph{et al.}, 2015~\cite{2015ApJ...801...74G}. 

\section{Sensitivity analysis}

\subsection{SWGO}

The Southern Wide field-of-view Gamma-ray Observatory (SWGO) is a wide field-of-view gamma-ray instrument that will be sensitive to gamma-rays with energies from hundreds of GeV to hundreds of TeV. It will consist of a ground-based air shower detector array, located in South America. The candidate sites for the observatory range from Peru, with a latitude of -13 degrees, to Argentina, with a latitude of -24 degrees~\cite{SWGO:2021lic}. 

The results in this work consider the Strawman design as described in the \textit{Science Case white paper} for the SWGO~\cite{2019arXiv190208429A}. While newer ideas and technologies have been introduced to improve upon the initial design, simulated instrument response functions (IRFs) of the updated versions are still under development~\cite{2022icrc.confE..23H}. Because of the SWGO energy resolution and, particularly the bias (difference between true energy and reconstructed energy) towards lower energies, the limits were only calculated for for dark matter masses above a lower bound of 500 GeV.

\subsection{Likelihood Method}

In gamma-ray astronomy, it is usual to employ Li\&Ma-like likelihood in order to make statistical inference of parameters of the source parameter~\cite{1983ApJ...272..317L}. Having this in mind, the \textit{likelihood} function $L$ for a expected signal $S$, background $B$, and on-target count $N$ is defined as

\begin{equation}
     L =\frac{(B + S)^{N} e^{-(B + S)}}{N!} \, .
\end{equation}

Besides that, it is also important to consider that since the astrophysical factors (J-factor and D-factor) in dwarf galaxies are actually measured, we are able to consider its statistical uncertainties in the sensitivity calculation through a \textit{nuisance} parameter. The J-factor \textit{nuisance} parameter can be introduced to the \textit{likelihood} through a function of the type \cite{2015PhRvL.115w1301A, 2019PhRvD.100h3003H, 2020ChPhC..44h5001H}

\begin{equation}
     \label{eqn:jlike}
     \mathcal{J} = \frac{1}{\sqrt{2\pi}\sigma_j ln(10) \bar{J}}e^{-(log_{10}J-log_{10}\bar{J})^2/2\sigma_J ^2} \, ,
\end{equation}
where $\bar{J}$ is the measured value of J-factor with error $\sigma_J$ ($1\sigma$ Root Mean Square). This results in a log normal distribution with peak value $\bar{J}$ for each J-factor. A similar equation is used in the case of decaying dark matter and the D-factor.

Finally, a 2D-binned likelihood approach is used to increase the significance of results. This involves analyzing the signal and background in separate energy and spatial intervals (bins). For the case of Dwarf Galaxies, the spatial bins are usually the different galaxies. With that, the best value of $J$ is the one that maximizes the combined likelihood $\mathcal{L}$, for a given $\langle \sigma v \rangle$ and $m_{DM}$, 

\begin{equation}
\label{eqn:comblkl}
    \mathcal{L} = \prod_{i} \prod_{j} \mathcal{L}_{ij}  = \prod_{i} \prod_{j} L_{ij} \times \mathcal{J}_{j} \, ,
\end{equation}
for $i$ energy bins and $j$ spatial bins. By analyzing this likelihood function, constraints on $\langle \sigma v \rangle$ or $\tau$ are derived. This is done by comparing a hypothesis of excess signal with a null hypothesis, which represents the background and known signals. This is known as the log-likelihood ratio test and it is given by $TS = -2ln(\mathcal{L}/\mathcal{L}_0)$, where $\mathcal{L}$ is the probability of alternative hypothesis (dark matter exists) and $\mathcal{L}_0$ is the probability of null hypothesis (dark matter do not exists / only background). The 95\% confidence level exclusion limits are established for values of $\langle \sigma v \rangle$ or $\tau$ for which $TS$ is greater than 2.71.

\section{Results and discussion}

\subsection{Sensitivity limits}

\begin{figure}
 \centering
 \includegraphics[width=\textwidth]{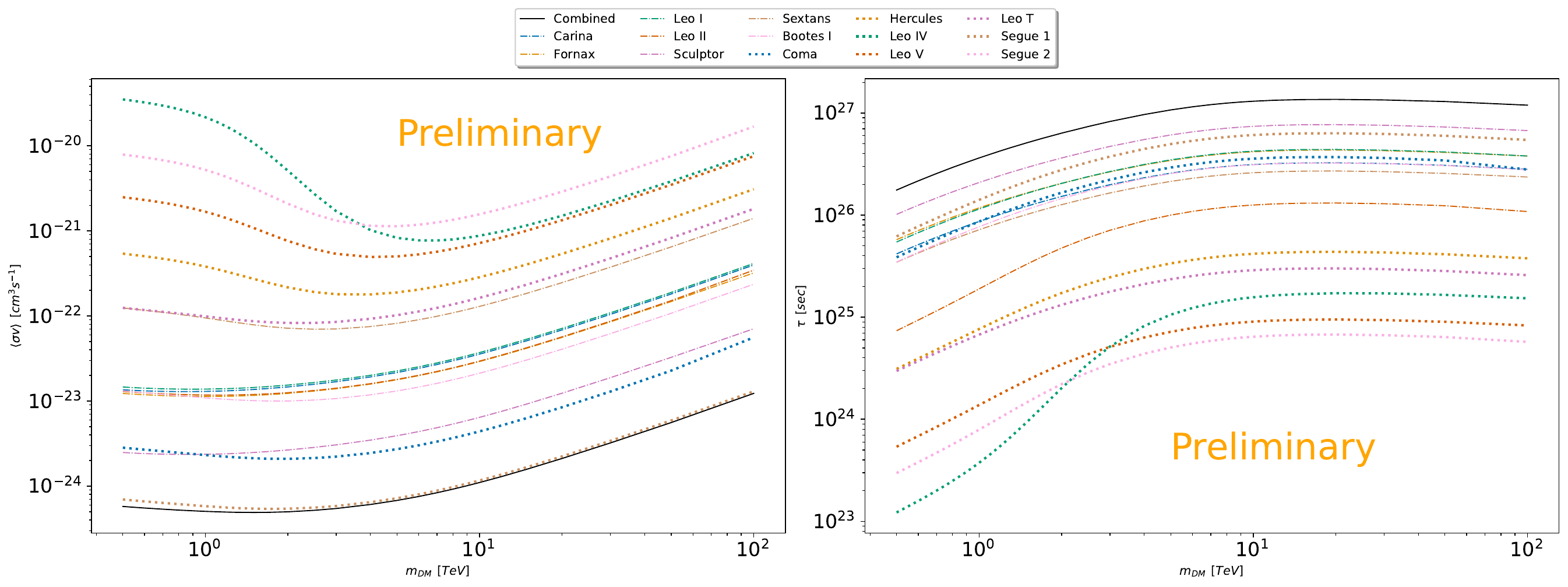}
 \caption{Sensitivity limits at 95\% confidence level for the annihilation cross-section and decay lifetime of dark matter annihilating/decaying into $\tau^+\tau^-$ as function of the dark matter particle mass. The limits for each dwarf galaxy are plotted for comparison with the combined limits (solid black line).}
 \label{fig_limits}
\end{figure}

Using the techniques shown thus far, we are able to establish 95\% confidence level limits for the annihilation cross-section and decay lifetime. Figure \ref{fig_limits} shows the individual and combined limits for different galaxies and dark matter particle masses ranging from 500 GeV to 100 TeV. These limits are determined for annihilation/decay with a 100\% branching ratio into quarks, leptons and bosons, in order to ensure model independence, but in Figure \ref{fig_limits} we present only the $\tau^+ \tau^-$ channel. 

In the case of dark matter particle of mass 1 TeV, the combined upper limits for the annihilation cross-sections reach the level of $\sim 5\times10^{-25}$ cm$^3$ s$^{-1}$ for the $\tau^+\tau^-$ channel. These limits are mainly dominated by Segue 1 as this is the dwarf galaxy with the highest J-factor from the fourteen in the range of the SWGO. Similarly, for a dark matter mass of 1 TeV the combined lower limits for the decay lifetimes reach $\sim 3\times10^{26}$ s for the $\tau^+\tau^-$ channel. In comparison to the annihilation limits, the decay limits are not dominated by any single dwarf and the usage of a combined analysis becomes very useful since it allows for a large gain in comparison to using any individual dwarf galaxy.

\subsection{Comparison with other experiments}

\begin{figure}
 \centering
 \includegraphics[width=\textwidth]{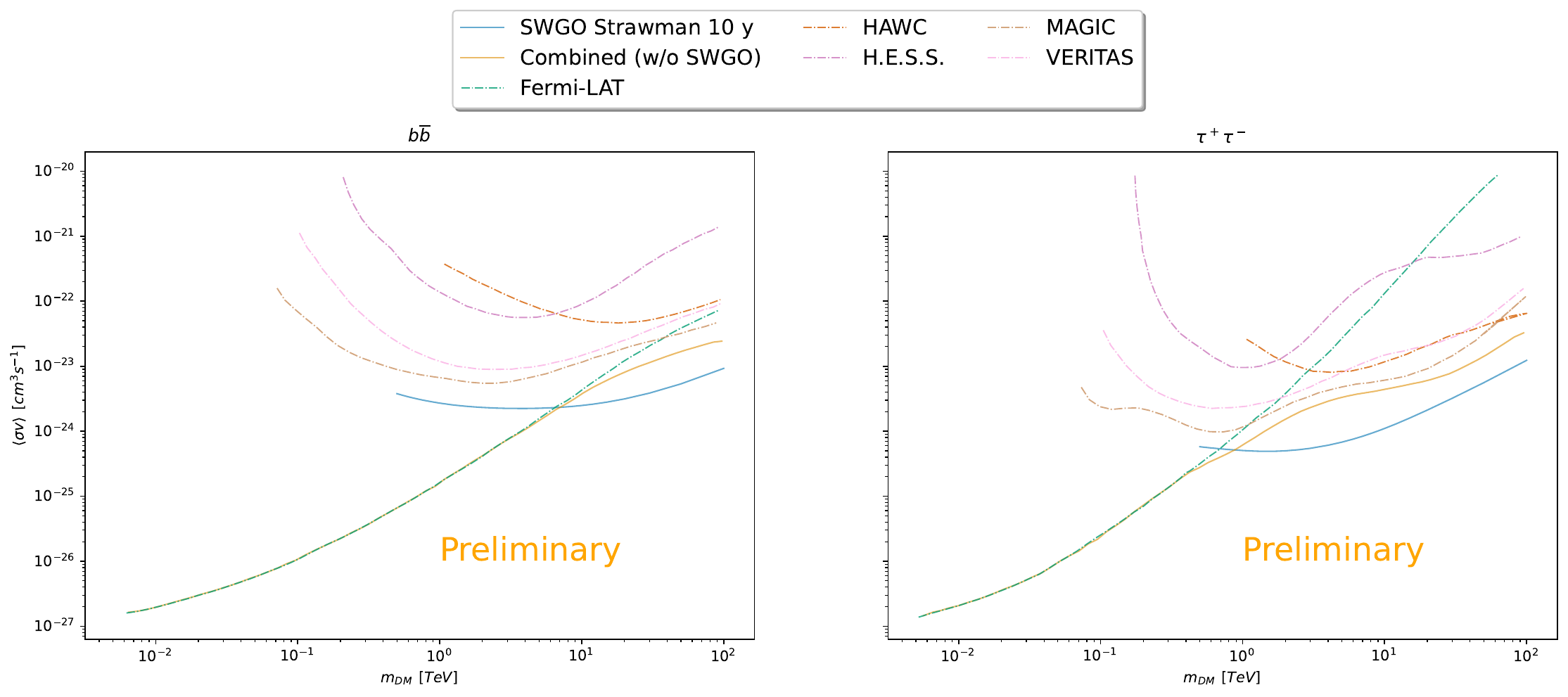}
 \caption{Comparison of the findings with the upper limits obtained by the combination for other five $\gamma$-ray experiments (Fermi-LAT, HAWC, H.E.S.S., MAGIC and VERITAS). The solid red line represents the combined results from the five $\gamma$-ray experiments while the solid blue line represents the curve for the SWGO.}
 \label{fig_combinedcomparison}
\end{figure}

It is also important to compare the SWGO capabilities with other experiments. We compare the obtained results to the combined limits obtained by the Fermi-LAT, HAWC, H.E.S.S., MAGIC, and VERITAS collaborations~\cite{2022icrc.confE.528A}. This collaboration between five different $\gamma$-ray instruments established annihilation limits for the combination of 20 dSphs observations. The comparison is shown in Figure \ref{fig_combinedcomparison}. The improvements of the limits by SWGO are notable in the multi-TeV range of dark matter masses, surpassing the capabilities of existing $\gamma$-ray instruments.

\section{Acknowledgments}

This work has made use of the IRFs from the SWGO Collaboration and has been approved by SWGO Collaboration's internal review.
This work is supported by FAPESP, grants 2021/09744-9, 2019/14893-3, and 2021/01089-1.
AV is also supported by CNPq grant 314955/2021-6.

%
%
%

\end{document}